\documentclass[amsmath,nofootinbib,notitlepage,prd,superscriptaddress,twocolumn]{revtex4-2}

\usepackage[T1]{fontenc}
\usepackage[usenames,dvipsnames]{xcolor}
\usepackage{aas_macros,graphicx,hyperref,orcidlink}
\usepackage{amsfonts,amsmath}
\usepackage{graphicx}
\usepackage{newtxtext,newtxmath}
\usepackage[normalem]{ulem}
\usepackage{cancel}
\usepackage{tikz}

\allowdisplaybreaks
\numberwithin{equation}{section}

\makeatletter
\renewcommand*{\@textcolor}[3]{%
  \protect\leavevmode
  \begingroup
    \color#1{#2}#3%
  \endgroup
}
\makeatother

\newcommand{\triright}{%
    \mbox{%
        \begin{tikzpicture}[baseline=-0.5ex, line cap=round]
            \draw[line width=0.12ex] (0,0) -- (-0.1, 0.1) 
                                     (0,0) -- (-0.1, -0.1) 
                                     (0,0) -- (0.1, 0.0);
        \end{tikzpicture}%
    }%
}

\newcommand{\pa}[1]{\left(#1\right)}
\newcommand{\pd}{\mathop{}\!\partial}

\begin{document}
\unitlength = 1mm
\setlength{\parskip}{1em}

\title{Polarization signatures of inspiraling hotspots around Kerr black holes}

\author{Pablo Ruales\,\orcidlink{0000-0002-3199-1025}}
\email{rualpm25@wfu.edu}
\affiliation{Department of Physics, Wake Forest University, Winston-Salem, North Carolina 27109, USA}

\author{Delilah E.~A. Gates\,\orcidlink{0000-0002-4882-2674}}
\affiliation{Center for Astrophysics $\arrowvert$ Harvard \& Smithsonian, 60 Garden Street, Cambridge, MA 02138, USA}
\affiliation{Black Hole Initiative at Harvard University, 20 Garden Street, Cambridge, MA 02138, USA}

\author{Alejandro C\'ardenas-Avenda\~no\,\orcidlink{0000-0001-9528-1826}} 
\affiliation{Department of Physics, Wake Forest University, Winston-Salem, North Carolina 27109, USA}

\begin{abstract}
Polarimetric interferometry is a powerful tool for probing both black hole accretion physics and the background spacetime. Current models aimed at explaining the observed multiwavelength flares in Sgr~A* often assume hotspots moving on geodesic, Keplerian orbits. In many scenarios, though, a hotspot may instead follow an inspiraling trajectory, potentially transitioning into a plunge toward the black hole. In this work, we present a general framework to simulate the polarized emission from generic equatorial inspiraling hotspots in Kerr spacetime using a parametric four-velocity profile. This parametrization defines a continuous family of flows, ranging from Cunningham’s disk model (fixed radius orbits outside the innermost stable circular orbit and plunging motion within the innermost stable circular orbit) to purely radial motion, thereby extending the standard assumptions. Within this framework, we show that inspiral motion produces a distinctive observational signature: a precessing, unwinding evolution of the polarimetric Stokes $Q$--$U$ looping pattern, in sharp contrast with the closed $Q$--$U$ loops associated with stable orbits at a fixed radius. We then explore how the morphology of these signatures depends on black hole spin, observer inclination, and magnetic-field configuration. The presented model can be applied to current and near-future interferometric observations of linear polarization, offering a new avenue to probe the physics of matter spiraling inward and the relativistic velocities of plunging plasma.
\end{abstract}

\maketitle

\vspace{-2em}

\section{Introduction}

The landmark images of the supermassive black holes M87* and Sgr~A* produced by the Event Horizon Telescope (EHT) collaboration have unveiled a rich phenomenology in the strong-gravity regime that remains to be fully characterized~\cite{EventHorizonTelescope:2022wkp}. In these extreme environments, matter in the inner accretion flow is accelerated to relativistic velocities, giving rise to complex interactions. One interesting consequence of these dynamics is the occurrence of flares, which are often interpreted as the formation of localized, over-dense regions, or ``hotspots,'' that emit intensely against the background flow. These hotspots might be produced by magnetic flux tubes or plasmoids~\cite{Levis:2023tpb,Ball:2020jup,Jiang:2025huk} generated via magnetic reconnection~\cite{Jiang:2025huk,Porth:2020txf,Ripperda:2021zpn}, or even by the tidal stripping of orbiting sub-stellar objects~\cite{Seoane:2025hlg}. For Sgr~A*, such events have been observed in millimeter~\cite{Wielgus:2022heh,EventHorizonTelescope:2022ago}, infrared~\cite{GRAVITY:2023avo,vonFellenberg:2025vrk}, and x-ray wavelengths~\cite{Mossoux:2020ddc}. 

Because directly resolving the coupled plasma physics and general relativistic effects governing these processes is extremely challenging, linear polarization provides a powerful diagnostic that probes both the magnetic-field structure and the spacetime geometry. Unlike total intensity, which is a scalar quantity, polarization carries vector information. Therefore, one can use the electric vector position angle (EVPA) observed at infinity to study the dynamics. The EVPA is directly related to the magnetic-field orientation in the emitting material’s rest frame and the geometry of spacetime. Hence, it is crucial to characterize how these polarimetric signatures are influenced by the competing effects of magnetic-field geometry and the strong-gravity of the spacetime~\cite{Broderick:2005my,Vincent:2023sbw}.

The Stokes parameters $Q$ and $U$ describe the polarized intensity on the observer's screen, and are directly related to the EVPA and the total polarized intensity. The observed flux of the signal is modulated by a redshift factor $g$ that accounts for Doppler boosting and gravitational redshift. In millimeter and near-infrared light curves of Sgr~A*~\cite{Wielgus:2022heh,GRAVITY:2023avo}, the Stokes parameters form loops, the so-called $Q$--$U$ loops. Their morphology can resemble a lima\c{c}on-like curve, with a prominent outer loop accompanied by a smaller inner loop, although in other cases the observed patterns can be highly elongated and the inner loop may be absent.

Hotspot models, which date back decades~\cite{1992A&A...257..531K}, have typically been applied to polarimetric observations under the assumption of stable circular trajectories outside the innermost stable circular orbit (ISCO)~\cite{Broderick:2005jj,Broderick:2005my,Gelles:2021kti,Wielgus:2022heh,Rosa:2025pqp,Yfantis:2024eab}. In Ref.~\cite{Wielgus:2022heh}, for example, a radio dataset of Sgr~A* was modeled with an orbital timescale of $t \sim 1.5$~hrs using a perpetual hotspot on a Keplerian orbit at $r = 11\,M$ (in geometric units), consistent with other studies~\cite{GRAVITY:2020lpa,GRAVITY:2023avo,Levis:2023tpb,Yfantis:2023wsp} that adopt radii in the range $r \sim 8 \,$--$ \,11\,M$. At these radii, the polarimetric signal often shows only a weak dependence on black hole spin or on the detailed magnetic-field configuration. While such models successfully reproduce the primary $Q$--$U$ loop~\cite{Wielgus:2022heh,Yfantis:2023wsp}, portions of the observed evolution fall outside what a perpetual hotspot at fixed radius can naturally produce. 

An inspiraling hotspot is expected to exhibit an intrinsically time-dependent polarimetric signature, providing a natural extension beyond current fits of the primary loop~\cite{Wielgus:2022heh,Yfantis:2023wsp}. In particular, the evolving $Q$--$U$ morphology found in the plunging region between the ISCO and the event horizon, as shown in Ref.~\cite{Chen:2024jkm}, captures the qualitative behavior expected whenever the emitting source undergoes systematic inward motion: the emergence of interior loops in the $Q$--$U$ plane and a systematic inward drift of the polarimetric centroid. 

Naturally, plunging hotspots that form near the ISCO and hotspots orbiting at larger radii probe different dynamical regimes: large-radius orbits are appropriate for slower variability originating farther out, whereas a plunging model becomes relevant when a primary loop is observed on significantly faster timescales. Moreover, because hotspots are generated by complex plasma physics, they need not move with strictly Keplerian velocities and may begin complex orbital motion before reaching the ISCO~\cite{Matsumoto:2020wul,Porth:2020txf,vonFellenberg:2025vrk}, particularly in an advection-dominated accretion flow such as that in Sgr~A*~\cite{Narayan:1994is}. Observations such as those studied in Ref.~\cite{Wielgus:2022heh} therefore motivate the need for a model that can handle generic inspiral motion, not only geodesic plunges within the ISCO. Indeed, plasmoid structures seen in general relativistic magnetohydrodynamic simulations suggest that a hotspot can originate in the accretion flow and be observed at mm wavelengths as it inspirals toward the event horizon~\cite{Zamaninasab:2009df,Emami:2022ydq}.

In this work, we take a step toward generic inspiral motion, not restricted to the region within the ISCO, by parameterizing the hotspot four-velocity. We adopt a prescription that spans trajectories from purely infalling motion to Cunningham's geodesic flow~\cite{Pu:2016qak,EventHorizonTelescope:2020eky,Cardenas-Avendano:2022csp}. This approach introduces two parameterizations: $(\beta_r, \xi)$ and $(\beta_r, \beta_\phi)$. The parameter $\xi$ sets the ratio of the fluid angular momentum to its Keplerian value, while $\beta_r$ and $\beta_\phi$ control the radial and azimuthal components of the motion by smoothly interpolating between the sub-Keplerian prescription and purely radial infall ($\xi$ and $\beta_\phi$ are not independent). Our prescription generalizes both the geodesic plunging motion considered in Ref.~\cite{Chen:2024jkm} and the fixed radius motion with a sub-Keplerian parameter studied in Refs.~\cite{Vos:2022yij,Levis:2023tpb}.

By specifying the hotspot dynamics and the magnetic-field geometry independently, we systematically characterize the distinctive morphologies that arise when orbital motion is coupled to radial infall motion, producing inspiraling trajectories. The model takes as an input the local magnetic-field configuration in the comoving frame, and is therefore flexible enough to accommodate a wide range of magnetic-field prescriptions. The detailed morphology and relevant timescales ultimately depend on the source's four-velocity, and therefore on the specific inspiral profile of the emitting region.

The structure of the paper is as follows. In Sec.~\ref{sec:theory}, we provide the theoretical background required to compute the $Q$--$U$ loops. In Sec.~\ref{sec:inspiraling_hotspots}, we present the dynamical model for generic inspiraling hotspots in Kerr geometry. We conclude in Sec.~\ref{sec:discussion} with a discussion of our results and future outlook. Throughout this work, we adopt geometric units ($G=1=c$).

\section{The theoretical minimum of the \textbf{$Q$--$U$} Loops}\label{sec:theory}

\begin{figure*}
    \includegraphics[width=1\textwidth]{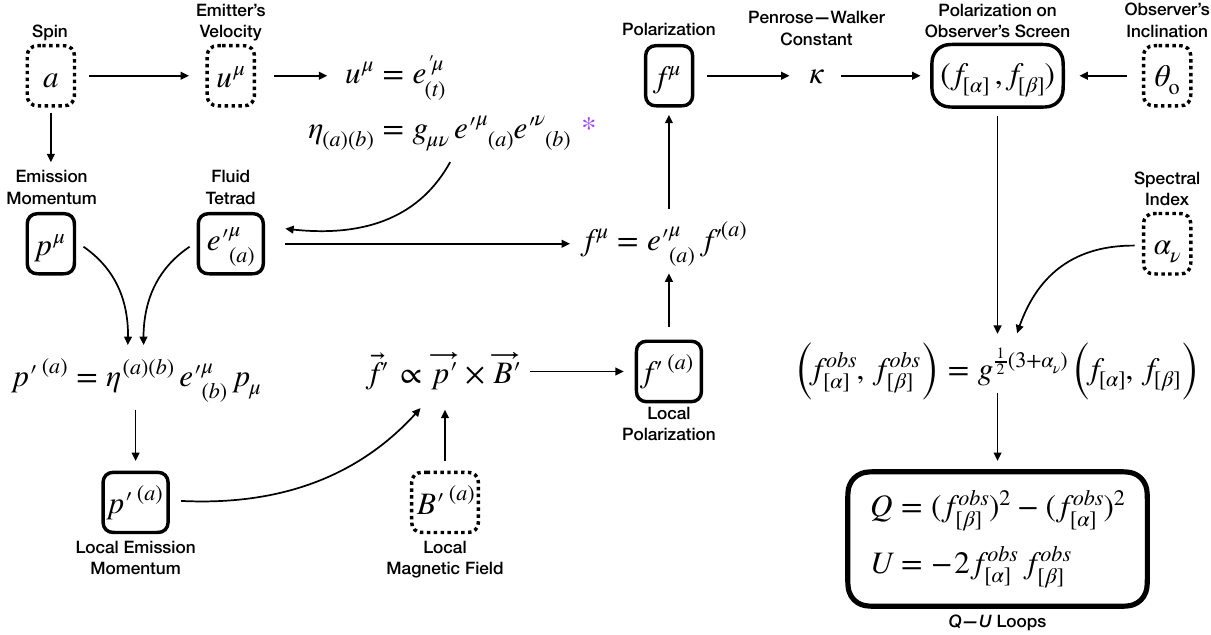}
    \caption{This diagram summarizes the pipeline used to compute the Stokes $Q$ and $U$ parameters. It traces how the photon momentum ($p^\mu$), defined in the background geometry, is projected into the fluid frame, and how the synchrotron polarization is then transformed back to the background. The polarization is propagated to the observer’s screen via the Penrose--Walker constant (which is invariant along null geodesics) and then rescaled by a redshift factor to obtain the observed linear polarization in terms of the Stokes $Q$ and $U$ parameters. Dashed boxes indicate model inputs, while solid-line boxes denote quantities computed directly using the equations presented in Sec.~\ref{sec:theory}. The calculation used to obtain the boosted set of orthonormal vectors and one-forms (highlighted in the figure with the symbol \textcolor{Purple}{*}) is detailed in Fig.~\ref{fig:transformation}.}
    \label{fig:diagram}
\end{figure*}

\begin{figure}
    \includegraphics[width=0.9\columnwidth]{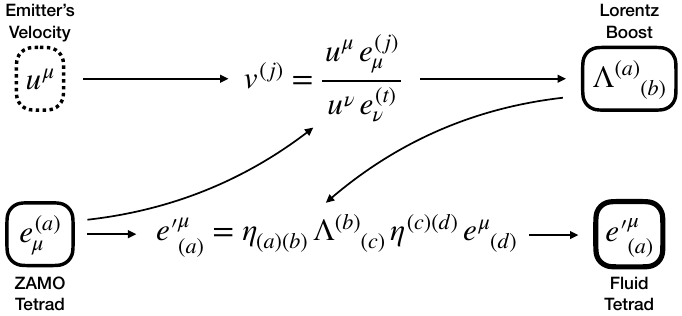}
    \caption{This diagram shows the two orthonormal bases of vectors and one-forms used to transform quantities from the global geometry to either the ZAMO frame ($e_{\mu}$) or the fluid frame ($e'_{\mu}$). Boxes with dashed-line frames indicate model inputs, while boxes with solid-line frames denote quantities computed analytically.}
    \label{fig:transformation}
\end{figure}

Computing the $Q$--$U$ loops requires an astrophysical model for the emission and motion of gas in the disk, as well as a background spacetime geometry, which we take here to be Kerr. The overall procedure for computing these Stokes parameters on the observer’s screen is illustrated in Figs.~\ref{fig:diagram} and~\ref{fig:transformation}: dashed-line boxes denote input parameters (the background metric and the astrophysical model), while solid-line boxes denote quantities computed in the pipeline.

In this Section, we describe each step of the procedure in detail, keeping the presentation as general as possible within the Kerr geometry. Although the key elements of this methodology appear in previous works (e.g., Refs.~\cite{Gelles:2021kti,EventHorizonTelescope:2021btj}), they are often dispersed across different conventions and levels of detail. We therefore collect and present the theoretical ingredients in a unified notation, with sufficient intermediate steps to make the full pipeline reproducible. Importantly, we do so without specifying the emitter four-velocity or the details of the astrophysical model. Readers interested primarily in the inspiraling model may skip to Sec.~\ref{sec:inspiraling_hotspots}.

\subsection{The Fluid Reference Frame}
\label{sec:ZAMO_geometry}

Because the disk astrophysics is typically described locally, and therefore in an approximately flat manifold, we must connect the radiating source’s local description to the global Kerr black hole background (hereafter, the Kerr frame), characterized by mass $M$ and angular momentum $J=Ma$, which we express in Boyer--Lindquist coordinates. The essential mathematical tools for this task are tetrads: orthonormal basis vectors that define a local coordinate transformation and specify the frame of an object moving with four-velocity $u$. Mathematically, these tetrads are defined by
\begin{align}
    &e^\mu_{(t)}= u^\mu,\\
    \eta^{(a)(b)} &= g^{\mu\nu} \, e_\mu{}^{(a)} \, e_\nu{}^{(b)}, \label{eq:ortho_contra} \\[5pt]
    \eta_{(a)(b)} &= g_{\mu\nu} \, e^\mu{}_{(a)} \, e^\nu{}_{(b)}, \label{eq:ortho_cov}
\end{align}
where Latin indices $(a), (b), \ldots$ denote components in the new orthonormal frame, Greek indices $\mu, \nu, \ldots$ denote components in the coordinate Kerr frame, and $\eta_{(a)(b)} = \eta^{(a)(b)} = \mathrm{diag}(-1, 1, 1, 1)$ is the Minkowski metric of the local Lorentz frame (see, e.g., Ref.~\cite{Gates:2020sdh}). Equations (\ref{eq:ortho_contra}) and (\ref{eq:ortho_cov}) ensure that, at each spacetime point, the new frame obeys the laws of special relativity. The contravariant and covariant tetrads are defined, respectively, as,
\begin{align}
\mathbf{e}_{(a)} &= e^{\mu}{}_{(a)} \, \frac{\partial}{\partial x^{\mu}}, \\
\mathbf{e}^{(a)} &= e_{\mu}{}^{(a)} \, dx^{\mu}.
\end{align}
These tetrads allow tensor components to be transformed between the Kerr frame and a local orthonormal frame. For example, a covariant vector $h_{\mu}$ in the Kerr frame is transformed to its components $h^{(a)}$ in the local frame according to:
\begin{equation}
h^{(a)} = \eta^{(a)(b)} \, e^{\mu}{}_{(b)} \, h_{\mu}.
\label{eq:kerr_to_zamo}
\end{equation}
Next, we describe how to construct the orthonormal frame of the emitting material, $e'_{(a)}$, which we refer to as the ``fluid frame'' and denote with a prime. The procedure builds the fluid frame by relating it to the zero-angular-momentum observer (ZAMO) frame, as summarized in Fig.~\ref{fig:transformation}. Crucially, this construction provides a consistent way to project fields and particle motion defined in the fluid frame into the Kerr frame, using the ZAMO frame as an intermediate step.

Observers that co-rotate with the spacetime geometry are known as ZAMOs, whose defining property is that their angular momentum vanishes ($L = p_\phi = 0$), despite having a nonzero angular velocity. The contravariant tetrads defining the ZAMO frame are given by~\cite{Bardeen:1972fi},
\begin{equation}
    \begin{aligned}
        \mathbf{e}_{(t)} &= \left( \sqrt{\frac{A}{\Sigma \Delta }},\ 0,\ 0,\ \frac{2 a M r}{\sqrt{A \Sigma \Delta }} \right), \\[8pt]
        \mathbf{e}_{(r)} &= \left( 0,\ \sqrt{\frac{\Delta }{\Sigma }},\ 0,\ 0 \right), \\[8pt]
        \mathbf{e}_{(\theta)} &= \left( 0,\ 0,\ \frac{1}{\sqrt{\Sigma }},\ 0 \right), \\[8pt]
        \mathbf{e}_{(\phi)} &= \left( 0,\ 0,\ 0,\ \sqrt{\frac{\Sigma }{A }} \frac{1}{\sin{\theta}} \right),
    \end{aligned}
    \label{eq:zamo_tetrad_contravariant}
\end{equation}

where the standard Kerr metric functions in Boyer--Lindquist coordinates are $\Delta = r^2 + (aM)^2 - 2Mr$, $\Sigma = r^2 + (aM)^2 \cos^2 \theta$, and $A = (r^2 + (aM)^2)^2 - (aM)^2 \Delta \sin^2 \theta$. This choice of frame minimizes the effects of frame dragging: these local observers effectively ``rotate with the geometry,'' providing a locally flat, orthonormal basis in which Kerr-related quantities take a simplified form~\cite{Bardeen:1970vja,Bardeen:1972fi}. The corresponding angular velocity, set by frame dragging in the rotating spacetime, is $\omega(r,\theta) = -g_{t\phi}/g_{\phi\phi}$.

The ZAMOs see the emitting material with four-velocity $u$ as moving with three-velocity $\vec{\mathcal{V}}$, which by the standard prescription~\cite{Bardeen:1972fi} is,
\begin{equation}
\mathcal{V}^{(a)} = \frac{u^\mu e_\mu{}^{(a)}}{u^\nu e_\nu{}^{(t)}},
\label{eq:3velBardeen}
\end{equation}
where $e_\mu{}^{(a)}$ is the covariant tetrad, related to the contravariant tetrad $e^\mu{}_{(a)}$ (given in Eq.~\ref{eq:zamo_tetrad_contravariant}) by the Minkowski metric: $e_\mu{}^{(a)} = \eta^{(a)(b)} g_{\mu\nu} e^\nu{}_{(b)}$. This relation ensures the denominator $u^\nu e_\nu{}^{(t)}$ is positive for future-directed timelike motion, yielding a correct physical three-velocity. 

Then, vectors in the fluid frame and ZAMO frame must be related by a Lorentz boost,
\begin{equation}\label{eq:zamo_to_boosted}
h'^{(a)} = \Lambda^{(a)}{}_{(b)} \, h^{(b)},
\end{equation}
where \( \Lambda^{(a)}{}_{(b)}(\vec{\mathcal{V}}) \) is the Lorentz transformation matrix.
For a velocity three-vector in the equatorial plane $\vec{\mathcal{V}} = (\mathcal{V}_r, 0, \mathcal{V}_\phi)$ the Lorentz boost is given by,
\begin{equation}
\Lambda^{(a)}{}_{(b)}(\vec{\mathcal{V}}) = 
\begin{pmatrix}
 \gamma  & -\mathcal{V}_r \gamma  & 0 & -\mathcal{V}_\phi  \gamma  \\
 -\mathcal{V}_r \gamma  & 1 + \frac{\mathcal{V}_r^2 (\gamma -1)}{\mathcal{V}^2} & 0 & \frac{\mathcal{V}_r \mathcal{V}_\phi  (\gamma -1)}{\mathcal{V}^2} \\
 0 & 0 & 1 & 0 \\
 -\mathcal{V}_\phi \gamma  & \frac{\mathcal{V}_r \mathcal{V}_\phi  (\gamma -1)}{\mathcal{V}^2} & 0 & 1 + \frac{\mathcal{V}_\phi^2 (\gamma -1)}{\mathcal{V}^2} \\
\end{pmatrix},
\label{eq:lorentz_boost_matrix}
\end{equation}
where $\mathcal{V}^2 = \mathcal{V}_r^2 + \mathcal{V}_\phi^2$, and $\gamma = (1 - \mathcal{V}^2)^{-1/2}$ is the corresponding Lorentz factor.

Having constructed the Lorentz boost from the ZAMO to the fluid, we can effectively boost into a frame, which is co-moving with the fluid. Hence we obtain the tetrad for the fluid frame,
\begin{equation}
\begin{split}
    e'^{\mu}{}_{(a)} &= \Lambda_{(a)}{}^{(b)} \, e^{\mu}{}_{(b)}\\
    &= \eta_{(a)(d)} \, \Lambda^{(d)}{}_{(c)} \, \eta^{(c)(b)} \, e^{\mu}{}_{(b)}.
\end{split}
\label{eq:boosted_tetrad_contravariant}
\end{equation}
The corresponding covariant form of the fluid frame tetrad, necessary for transforming covariant objects, is given by the inverse transformation,
\begin{equation}
e'_{\mu}{}^{(a)} = \Lambda^{(a)}{}_{(b)} \, e_{\mu}{}^{(b)}.
\label{eq:boosted_tetrad_covariant}
\end{equation}

Finally, any vector is transformed from the fluid frame back to the global Kerr frame using the boosted tetrad,
\begin{equation}
h^{\mu} = e'^{\mu}{}_{(a)} \, h'^{(a)}.
\label{eq:boosted_to_kerr}
\end{equation}
This complete set of transformations, Eqs.~\eqref{eq:kerr_to_zamo}, \eqref{eq:zamo_to_boosted}, and \eqref{eq:boosted_to_kerr}, provides a consistent pipeline to map quantities from the global Kerr frame to a locally flat, comoving frame tied to the accretion flow, and vice versa.

\subsection{Linear Polarization}
\label{sec:polarization}

Because the relevant astrophysical processes are described locally in the fluid frame, we take the magnetic-field configuration $\vec{B'}$ as an input. The synchrotron polarization vector $\vec{f'}$ is perpendicular to both the particle momentum and the magnetic-field lines, and therefore satisfies~\cite{EventHorizonTelescope:2021btj},
\begin{equation}
\vec{f'} = \frac{\vec{p'} \times \vec{B'}}{|\vec{p'}|},
\label{eq:polarization_cross}
\end{equation}
where $\vec{p'}$, $\vec{B'}$, and $\vec{f'}$ are the spatial components $\left(r,\theta,\phi\right)$ of the momentum of the emitted photon, magnetic field, and polarization vectors in the fluid frame, respectively.

To compute the momentum $\vec{p'}$ in this frame, we begin with the covariant four-momentum $p_\mu$  of a photon on a geodesic orbit in the Kerr metric. At the emission location on the equatorial plane ($\theta = \pi/2$), the components are~\cite{Carter:1968rr},
\begin{equation}
\begin{aligned}
\frac{p_t}{E} &= -1, \\
\frac{p_r}{E} &= \pm_r \frac{\sqrt{\mathcal{R}(r)}}{\Delta}, \\
\frac{p_\theta}{E} &=\sqrt{\eta}, \\
\frac{p_\phi}{E} &=\lambda,
\end{aligned}
\label{eq:covariant_momentum}
\end{equation}
where $\lambda$ is the energy-rescaled axial angular momentum, $\eta$ is the energy-rescaled Carter constant, and $\mathcal{R}(r)$ is the radial potential as
\begin{equation}
\mathcal{R}(r) = \left(r^2 + (aM)^2 - a M \lambda\right)^2 - \Delta \left[\eta + (\lambda - aM)^2\right],
\label{eq:R_function}
\end{equation}
and the sign $\pm_r$ is determined by the radial direction of motion. The constants of motion $\lambda$ and $\eta$ are related to angular momentum $L_z$ and the Carter constant $Q$ by,
\begin{align}
    \lambda &= -\frac{p_\phi}{p_t} = \frac{L_z}{E},\\
    \eta &= \frac{p_\theta^2}{p_t^2} - (aM)^2 \cos^2 \theta + \lambda^2 \cot^2 \theta = \frac{Q}{E^2}.
\end{align}
Finally, as the photon trajectory does not depend on the energy, we set $E=1$ without loss of generality.

As described in the previous Section, the transformation of the covariant four-momentum $p_\mu$  from the global Kerr frame to its representation in the fluid frame is given by,
\begin{equation}
p'^{(a)} = \eta^{(a)(b)} e'^\mu{}_{(b)} p_\mu.
\label{eq:momentum_boost}
\end{equation}
In this expression, $e'^\mu{}_{(b)}$ are the contravariant fluid frame tetrads (Eq.~\ref{eq:boosted_tetrad_contravariant}). 
Once the polarization vector $\vec{f'}$ is calculated in the local frame via Eq.~\eqref{eq:polarization_cross}, it must be transformed back to the global Kerr frame to propagate the photons to a distant observer. This transformation is performed by replacing the local polarization $\vec{f'}$ with $\vec{h'}$ in Eq.~\ref{eq:boosted_to_kerr} to derive $f^\mu$.

Finally, we parallel transport the polarization along the photon trajectories from the source to the location of the observer. In the Kerr geometry, we can make use of the Penrose--Walker constant
$\kappa$~\cite{Walker:1970un},
\begin{equation}
\kappa = \kappa_1 + i \kappa_2 = r (\mathcal{P}_\mathcal{A} - i \mathcal{P}_\mathcal{B}),
\label{eq:penrose_walker}
\end{equation}
where the quantities $\mathcal{A}$ and $\mathcal{B}$ are constructed from the polarization and momentum components in the global frame~\cite{1977Natur.269..128C},
\begin{align*}
\mathcal{P}_\mathcal{A} &= (p^t f^r - p^r f^t) + a (p^r f^\phi - p^\phi f^r), \\
\mathcal{P}_\mathcal{B} &= (r^2 + a^2)(p^\phi f^\theta - p^\theta f^\phi) - a (p^t f^\theta - p^\theta f^t).
\label{eq:AB_components}
\end{align*}
The conserved value of $\kappa$ enables the direct computation of the observed polarized image.

\subsection{Observed Polarization}
\label{sec:observed_polarization}

To compute the observed polarization, we project the emitted radiation onto the celestial coordinates $(\alpha, \beta)$ of a distant observer's screen~\cite{Bardeen:1972fi}. These coordinates are functions of the photon's conserved quantities and the observer's inclination $\theta_{\rm{o}}$~\cite{Bardeen:1972fi},
\begin{equation}\label{eq:alphabeta}
    \alpha = -\frac{\lambda}{\sin \theta_{\rm{o}}}, \quad \beta = \pm_{\rm{o}} \sqrt{\eta + (aM)^2 \cos^2 \theta_{\rm{o}} - \lambda^2 \cot^2 \theta_{\rm{o}}},
\end{equation}
where the sign $\pm_{\rm{o}}$ distinguishes between the upper and lower halves of the observer's sky.

The polarization vector of a photon arriving at a point $(\alpha, \beta)$ on the screen is characterized by the EVPA, $\chi$. This angle can be determined by the Penrose--Walker constant~\eqref{eq:penrose_walker} and the screen coordinates ($\alpha,\beta$)~\cite{1994ApJ...427..718R},
\begin{equation}\label{eq:evpa}
    \chi = \arctan \left( \frac{\nu \kappa_1 - \beta \kappa_2}{\beta \kappa_1 + \nu \kappa_2} \right),
\end{equation}
where $\nu = -(\alpha + a \sin\theta_{\rm{o}})$. The components of the polarization vector $\vec{f}$ in the observer's screen coordinates (denoted with subscripts in hard brackets, i.e., $f_{[\alpha]}$) are given by,
\begin{equation}\label{eq:screen_pol_components}
    \left(f_{[\alpha]},f_{[\beta]}\right) = \frac{1}{\beta^2+\nu^2}\left( \beta \, \kappa_2 - \nu \, \kappa_1 , \beta \, \kappa_1 + \nu \, \kappa_2 \right).
\end{equation}
The observed polarization vector is then modified by two effects: the Doppler shift and gravitational redshift $g = \nu_{\rm{o}} / \nu_e$ (the ratio of observed to emitted frequency), and the spectral index $\alpha_\nu$ of the synchrotron radiation. The flux density of polarized emission scales as \cite{EventHorizonTelescope:2021btj}:
\begin{equation}\label{eq:f_obs_scaling}
    \left(f^{\text{obs}}_{[\alpha]},\ f^{\text{obs}}_{[\beta]}\right) = g^{\frac{1}{2}(3 + \alpha_\nu)} \left(f_{[\alpha]},\ f_{[\beta]}\right).
\end{equation}
This particular scaling is meant to account for the effects of aberration and the change in spectral energy distribution due to the relative motion between the source and the observer~\cite{Rybicki:2004hfl}. 

Finally, the Stokes parameters $Q$ and $U$ are~\cite{1994ApJ...427..718R,EventHorizonTelescope:2021btj,Himwich:2020msm,Gelles:2021kti}
\begin{align}
    Q &= \left(f^{\text{obs}}_{[\beta]}\right)^2 - \left(f^{\text{obs}}_{[\alpha]}\right)^2, \label{eq:Q} \\[5pt]
    U &= -2f^{\text{obs}}_{[\alpha]} f^{\text{obs}}_{[\beta]}. \label{eq:U}
\end{align}
The overall procedure for computing the $Q$--$U$ loops is shown in Figs.~\ref{fig:diagram} and~\ref{fig:transformation}.

\section{Inspiraling Hotspots}
\label{sec:inspiraling_hotspots}

The framework presented in the previous Section leaves the four-velocity of the equatorial, bound emitting hotspot unspecified. We now adopt a generic prescription for particle motion in the disk, which allows us to compute the polarized synchrotron emission from an inspiraling hotspot that can originate anywhere in the accretion flow and drift inward toward the black hole. We have implemented the procedure described in the previous Section in the open-source code \texttt{AART}~\cite{Cardenas-Avendano:2022csp}, an analytical ray-tracing code specialized to the Kerr geometry in the slow-light mode to account for the time-variable emission.

To model emission from inspiraling hotspots in the accretion flow, we adopt the velocity profile used in Refs.~\cite{Pu:2016qak,EventHorizonTelescope:2020eky,Cardenas-Avendano:2022csp}, which provides a general description of a non-geodesic inspiral in the equatorial plane. In this model, the contravariant four-velocity of the emitting particle is,
\begin{equation}\label{eq:four-velocity}
    \tilde{u}^\mu = \tilde{u}^t \left( \pd_t - \tilde{\iota} \pd_r + \tilde{\Omega} \pd_\phi \right),
\end{equation}
where $\tilde{\iota} = - \tilde{u}^r/\tilde{u}^t$, $\tilde{\Omega} = \tilde{u}^\phi/\tilde{u}^t$, are, respectively, the radial-infall and angular velocities, and
\begin{equation}
\label{eq:iota,omega}
\begin{aligned}
    \tilde{\ell}&=\xi\mathring{\ell},\\
    \tilde{u}^r&=\hat{u}^r+\pa{1-\beta_r}\pa{\bar{u}^r-\hat{u}^r},\\
    \tilde{\Omega}&=\hat{\Omega}+\pa{1-\beta_\phi}\pa{\bar{\Omega}-\hat{\Omega}}.\\
\end{aligned}
\end{equation}
Thus, this general flow is a linear superposition of circular motion and radial inflow: $\mathring{\ell}$ denotes the Keplerian specific angular momentum (i.e., the specific angular momentum for stable orbits at a fixed radius), $\hat{u}$ the four-velocity of the sub-Keplerian component, $\bar{u}$ the four-velocity of the radial inflow, $\hat{\Omega}$ the sub-Keplerian angular velocity, and $\bar{\Omega}$ the angular velocity of the radial inflow. The explicit expressions, which are lengthy, can be found in Appendix B of Ref.~\cite{Cardenas-Avendano:2022csp}.

This parametric four-velocity relates the radial-infall and angular velocities to the geodesic prescription of Cunningham~\cite{Cunningham:1975zz} through the parameters, $(\beta_r,\beta_\phi)$ or $(\beta_r,\xi)$, each ranging from $0$ to $1$. These parameters quantify the ``Keplerianity'' of the hotspot dynamics: $\xi$ sets the ratio of the fluid angular momentum outside the ISCO to its Keplerian value, resulting in a sub-Keplerian Cunningham-like flow; $\beta_r$ controls the radial motion by interpolating between the sub-Keplerian radial velocity and that of a particle dropped from rest at infinity; $\beta_\phi$ controls the azimuthal motion by interpolating between the sub-Keplerian angular velocity and that of a freely infalling particle with vanishing angular momentum. The parameters $\xi$ and $\beta_\phi$ are algebraically constrained and should not be varied independently. The parameter $\xi$ acts as a multiplicative factor rescaling the Keplerian covariant angular momentum, while $\beta_\phi$ rescales the Keplerian azimuthal angular velocity. Since these quantities correspond to different components of the four-velocity, $u_\phi$ and $u^\phi$, respectively, they are related through the metric. Including both parameters in the parametrization allows the formalism to accommodate different conventions for sub-Keplerian motion; see, for example, Refs.~\cite{Takahashi:2011dr,Penna:2013zga}, where sub-Keplerianity is defined in terms of the angular velocity and specific angular momentum, respectively.

By contrast, Cunningham's model~\cite{Cunningham:1975zz}, adopted in most previous studies, prescribes $\iota = 0$ outside the ISCO, yielding perpetual circular orbits with a fixed angular velocity $\Omega$. Plunging toward the event horizon occurs only within the plunging region $r < r_{\text{ISCO}}$, where Cunningham’s prescription determines both $\Omega$ and a non-zero radial velocity $\iota$ by conserving the energy and angular momentum of a particle crossing the ISCO.

This non-geodesic prescription can lead to a broad range of plunging times. For instance, a hotspot starting at $r = 11\,M$, i.e., the inferred location in Ref.~\cite{Wielgus:2022heh} for a perpetual hotspot, with $\beta_r = 0.90$ and $\beta_\phi = 0.98$, would reach the horizon in $t \sim 249\,M$ for $a = 0.94$ and $t \sim 187\,M$ for $a = 0$ (corresponding to $\sim 88$ minutes and $\sim 66$ minutes for Sgr~A*, respectively, assuming $M = 4.3 \times 10^6\,M_\odot$~\cite{EventHorizonTelescope:2022wkp}). In contrast, for $\beta_r = 0.98$ and $\beta_\phi = 0.98$, the hotspot would reach the horizon after $t \sim 1142\,M$ for $a = 0.94$ and $t \sim 769\,M$ for $a = 0$ (corresponding to $\sim 6.7$ hrs and $\sim 4.5$ hrs for Sgr~A*, respectively, assuming the same mass).

Having prescribed a general four-velocity~\eqref{eq:four-velocity}, the projection formula~\eqref{eq:3velBardeen} yields the physical three-velocity components of the hotspot in the ZAMO frame, $\vec{\mathcal{V}} = (\mathcal{V}^{(r)}, \mathcal{V}^{(\theta)}, \mathcal{V}^{(\phi)})$. This formalism provides a complete characterization of the local comoving frame for a hotspot on an inspiraling trajectory, enabling us to compute its emission properties as it moves toward the event horizon.

\begin{figure*}[]
    \centering
    \includegraphics[width=0.8\textwidth]{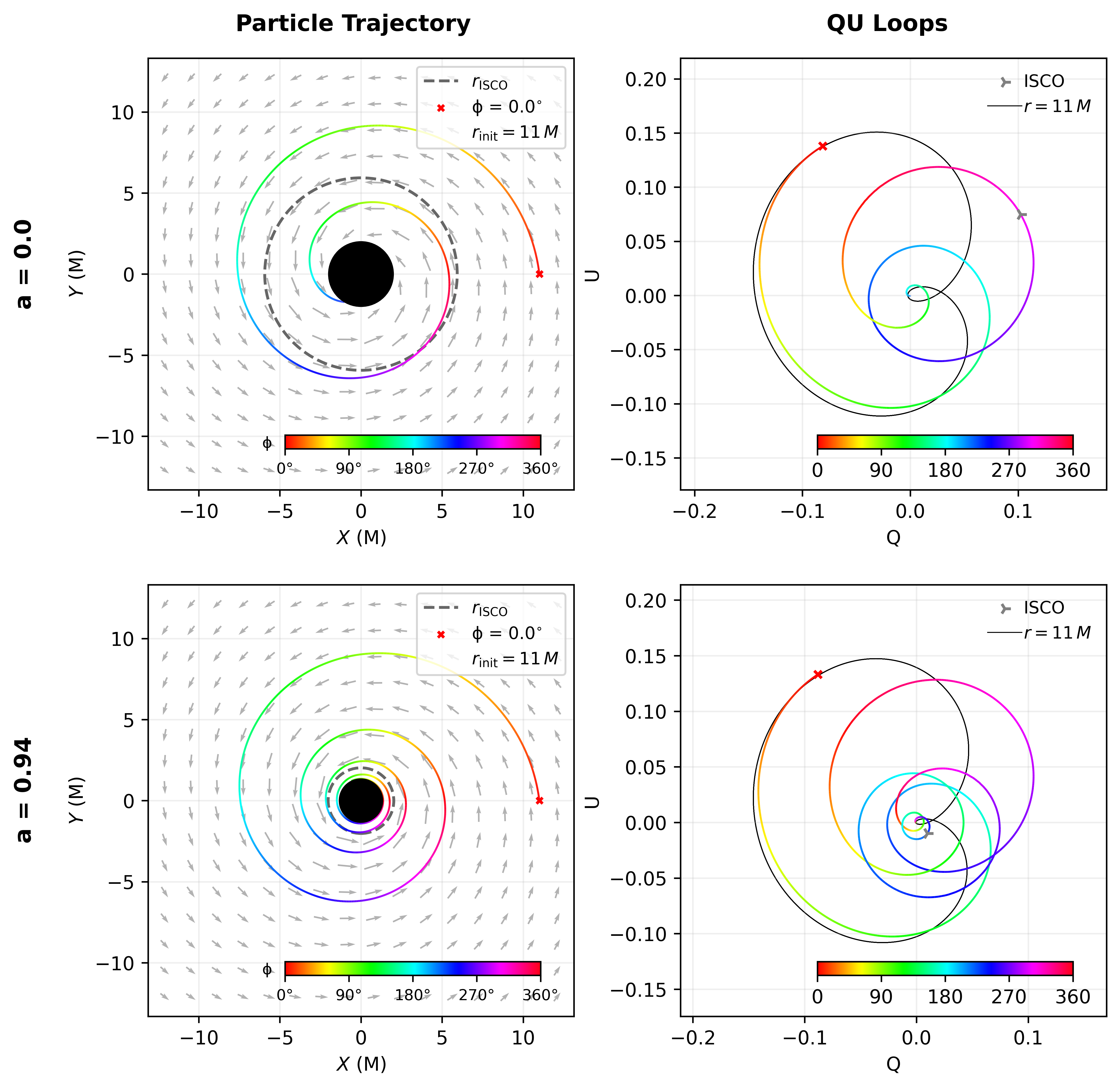}
    \caption{Inspiraling hotspot. The first column shows the hotspot trajectory, starting at $r = 11\,M$, around the source (top: $a = 0$; bottom: $a = 0.94$). The arrows indicate the equatorial velocity field vectors, which represent the velocity parametrization given in Eqs.~\ref{eq:iota,omega}. The second column shows the corresponding linear polarization of the hotspot emission on the observer’s screen (located in the upper hemisphere of the coordinate system at $\phi = 0$), represented in the $Q$--$U$ plane in geometrical units. The black line denotes a perpetual hotspot at the initial radius, and \protect\triright{} marks the ISCO crossing. In both examples, the local magnetic field is vertical, $\vec{B} = (0.0,1.0,0.0)$, and the observer is located at $\theta_{\rm{o}} = 20^{\circ}$. The inspiral is allowed to proceed all the way to the event horizon, with four-velocity (Eq.~\ref{eq:four-velocity}) parameters $\beta_r = 0.90$ and $\beta_\phi = 0.98$.}
    \label{fig:trajectories_new}
\end{figure*}

We now assess how the fundamental model parameters---black hole spin ($a$), observer inclination ($\theta_{\rm{o}}$), local magnetic-field configuration ($\vec{B}$), and spectral index ($\alpha_\nu$)---shape the morphology of the resulting polarimetric loops. We interpret the dynamics using the trajectories in Fig.~\ref{fig:trajectories_new}, which are later used to compute the polarization signatures in the $Q$--$U$ plane.

\begin{figure*}[]
    \centering
    \includegraphics[width=0.75\textwidth]{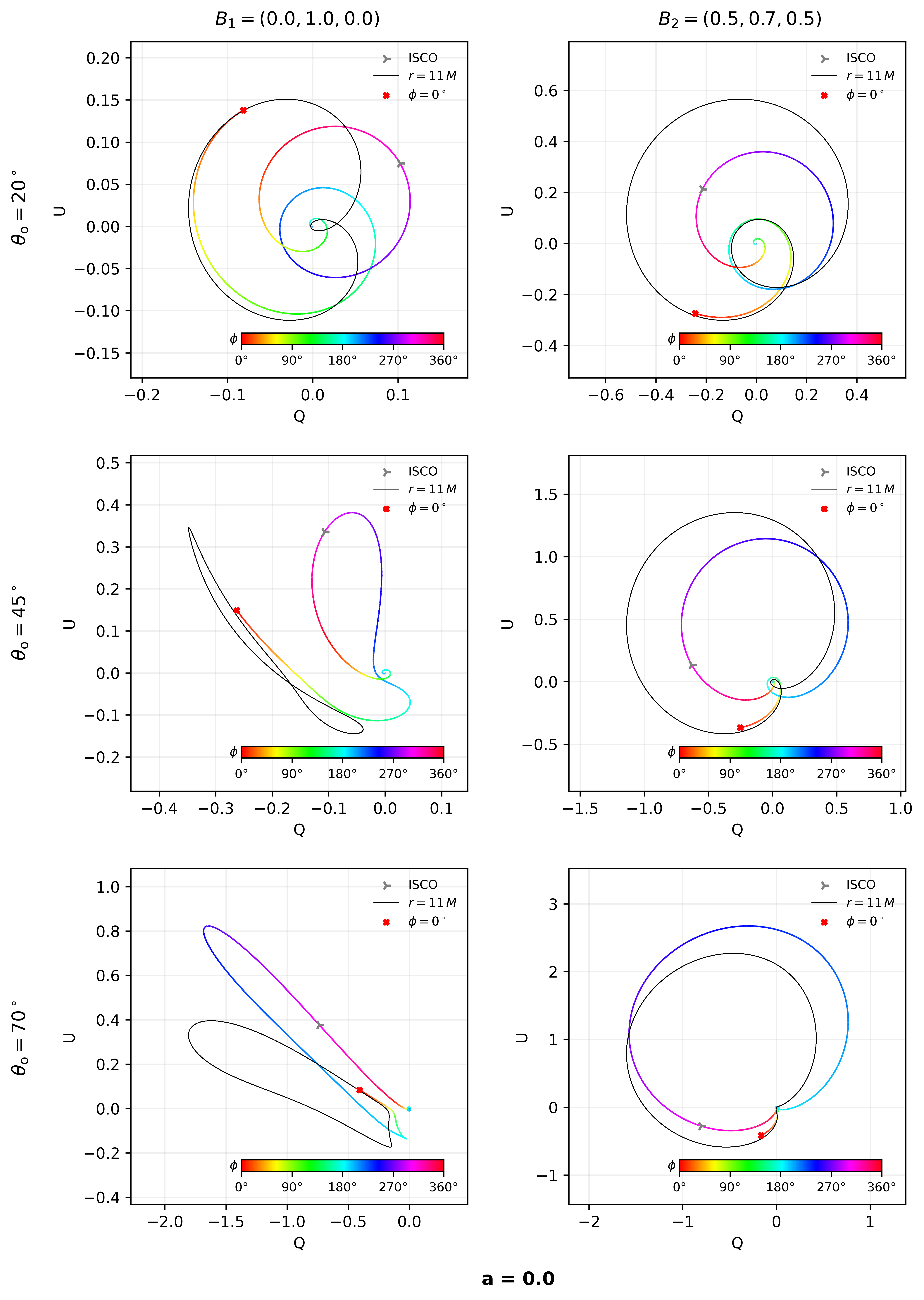}
    \caption{Polarimetric $Q$--$U$ loops for a hotspot inspiraling into a non-spinning black hole ($a = 0$). The columns vary the magnetic field configuration: purely vertical ($\vec{B}_1 = (0.0,1.0,0.0)$, left) and predominantly vertical with non-axisymmetric components ($\vec{B}_2 = (0.5,0.7,0.5)$, right). The rows vary the observer inclination ($\theta_o = 20^\circ, 45^\circ, 70^\circ$). The colored trajectory maps the instantaneous hotspot azimuth $\phi$; the black line denotes a stable circular orbit at $r = 11\,M$ for reference, and \protect\triright{} marks the ISCO crossing. The main features are as follows: (1) for $\vec{B}_1$, the loops are nearly circular at low inclination but become elongated as $\theta{\rm{o}}$ increases, whereas for $\vec{B}_2$ they remain nearly circular at all inclinations; (2) a clear inner secondary loop is present at low $\theta{\rm{o}}$ but disappears at high inclination; (3) the loop amplitude increases with $\theta{\rm{o}}$; (4) at low inclination, all orbital phases ($\phi = 0^\circ\text{--}360^\circ$) are visible, whereas at high $\theta{\rm{o}}$ emission from the receding phase ($\phi \sim 0^\circ\text{--}180^\circ$) is suppressed by gravitational redshift; and (5) the primary effect of modifying the $\vec{B}$-field configuration is to rotate the loop in the $Q$--$U$ plane and significantly alter its shape, suggesting that astrophysical effects dominate over gravitational ones during the polarization evolution. The four-velocity parameters in Eq.~\ref{eq:four-velocity} are $\beta_r = 0.90$ and $\beta_\phi = 0.98$.}
    \label{fig:low_spin_QU}
\end{figure*}

\begin{figure*}[]
    \centering
    \includegraphics[width=0.75\textwidth]{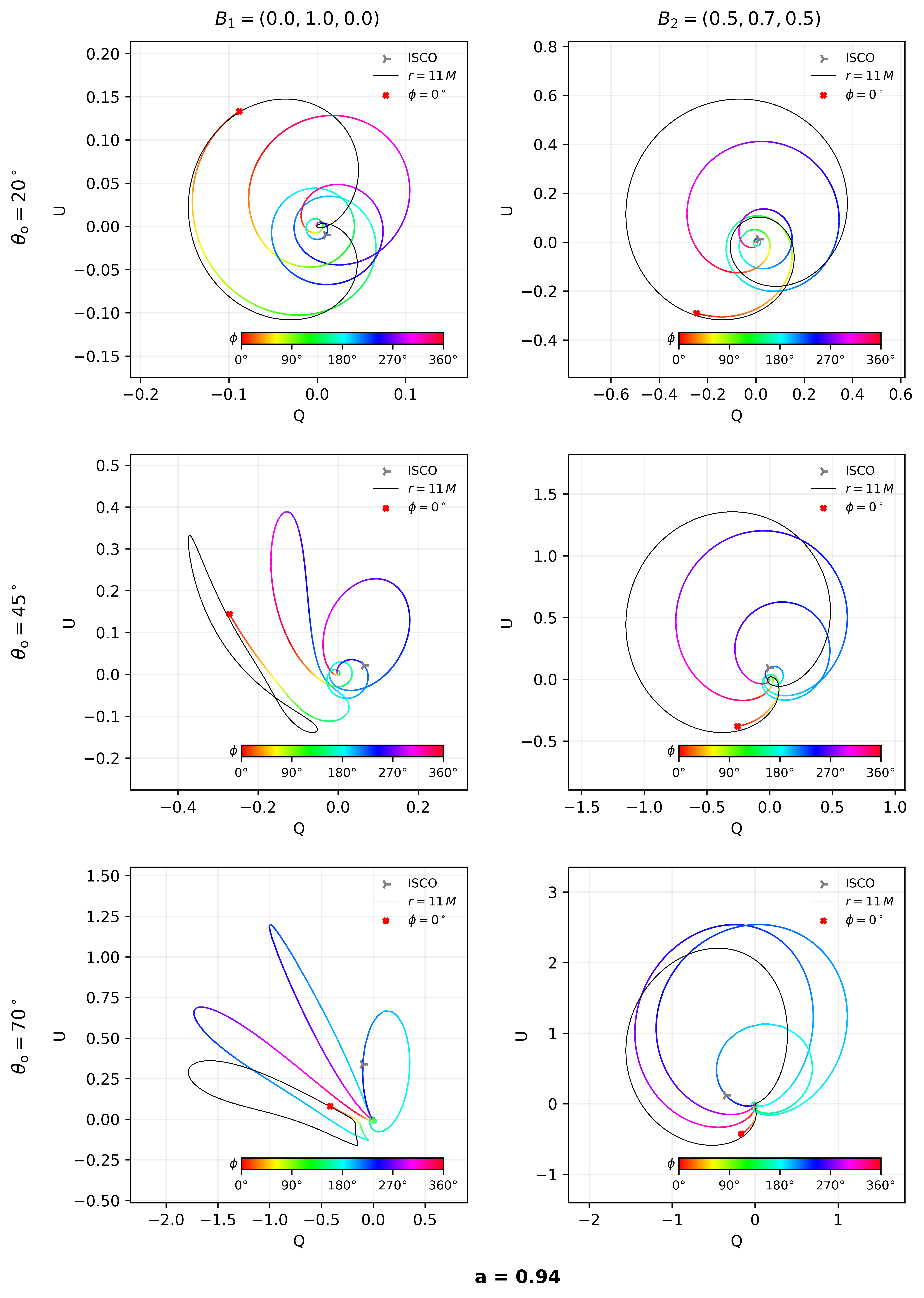}
    \caption{
    Polarimetric $Q$--$U$ loops for a hotspot inspiraling into a rapidly spinning black hole ($a=0.94$). Columns vary the magnetic field configuration: purely vertical ($\vec{B}_1=(0.0,1.0,0.0)$, left) and predominantly vertical with non-axisymmetric components ($\vec{B}_2=(0.5,0.7,0.5)$, right). Rows vary the observer inclination ($\theta_{\rm{o}} = 20^\circ, 45^\circ, 70^\circ$). The colored trajectory maps the instantaneous hotspot azimuth $\phi$; the black line is a stable circular orbit at $r=11\,M$ for reference, and \protect\triright{} shows the ISCO crossing. The key features that we can observe are: (1) all loops for $\vec{B}_2$ remain nearly circular, with only slight elongation at the highest inclination, while the $\vec{B}_1$ case shows elongation for increasing $\theta_{\rm{o}}$; (2) the color map becomes increasingly biased to $\phi \sim 180^\circ\text{-- }360^\circ$ (approaching phase) as the inclination increases, demonstrating extreme Doppler boosting that renders the receding phase nearly undetectable; and (3) changing the $\vec{B}$ field configuration results in a rotation of the loop in the $Q$--$U$ plane, evidenced in the low inclination case. The parameters for the four-velocity (Eq.~\ref{eq:four-velocity}) are $\beta_r = 0.90$ and $\beta_\phi = 0.98$.}
    \label{fig:high_spin_QU}
\end{figure*}

\begin{figure*}[]
    \centering
    \includegraphics[width=0.75\textwidth]{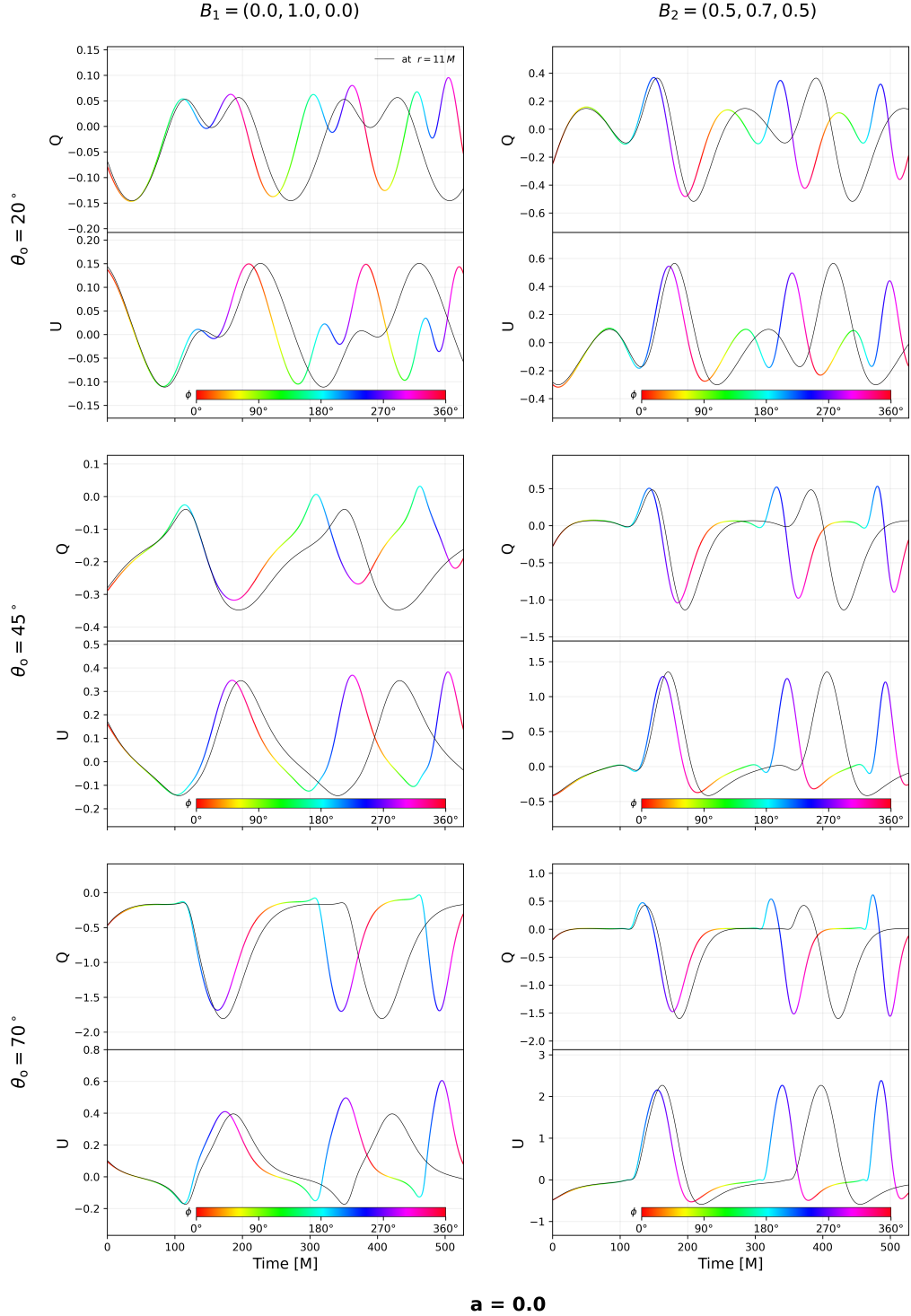}
    \caption{
    Time evolution of Stokes $Q$ (top row) and $U$ (bottom row), in arbitrary units, for a hotspot inspiraling into a non-spinning black hole ($a= 0$), which originates at $r = 11\,M$. Time is in gravitational units ($M$); for Sgr~A* ($M=4.3\times10^6\,M_\odot$),  $100\,M \approx 35$ minutes. Columns vary the magnetic field configuration, rows the observer's inclination. The colored trajectory (color indicates hotspot's azimuth $\phi$) shows the inspiraling case, while the black line shows a stable circular orbit at the initial radius, for reference, and \protect\triright{} shows the ISCO crossing. The parameters for the four-velocity (Eq.~\ref{eq:four-velocity}) are $\beta_r = 0.90$ and $\beta_\phi = 0.98$.}
    \label{fig:UvTandQvT_0spin}
\end{figure*}

\begin{figure*}[]
    \centering
    \includegraphics[width=0.75\textwidth]{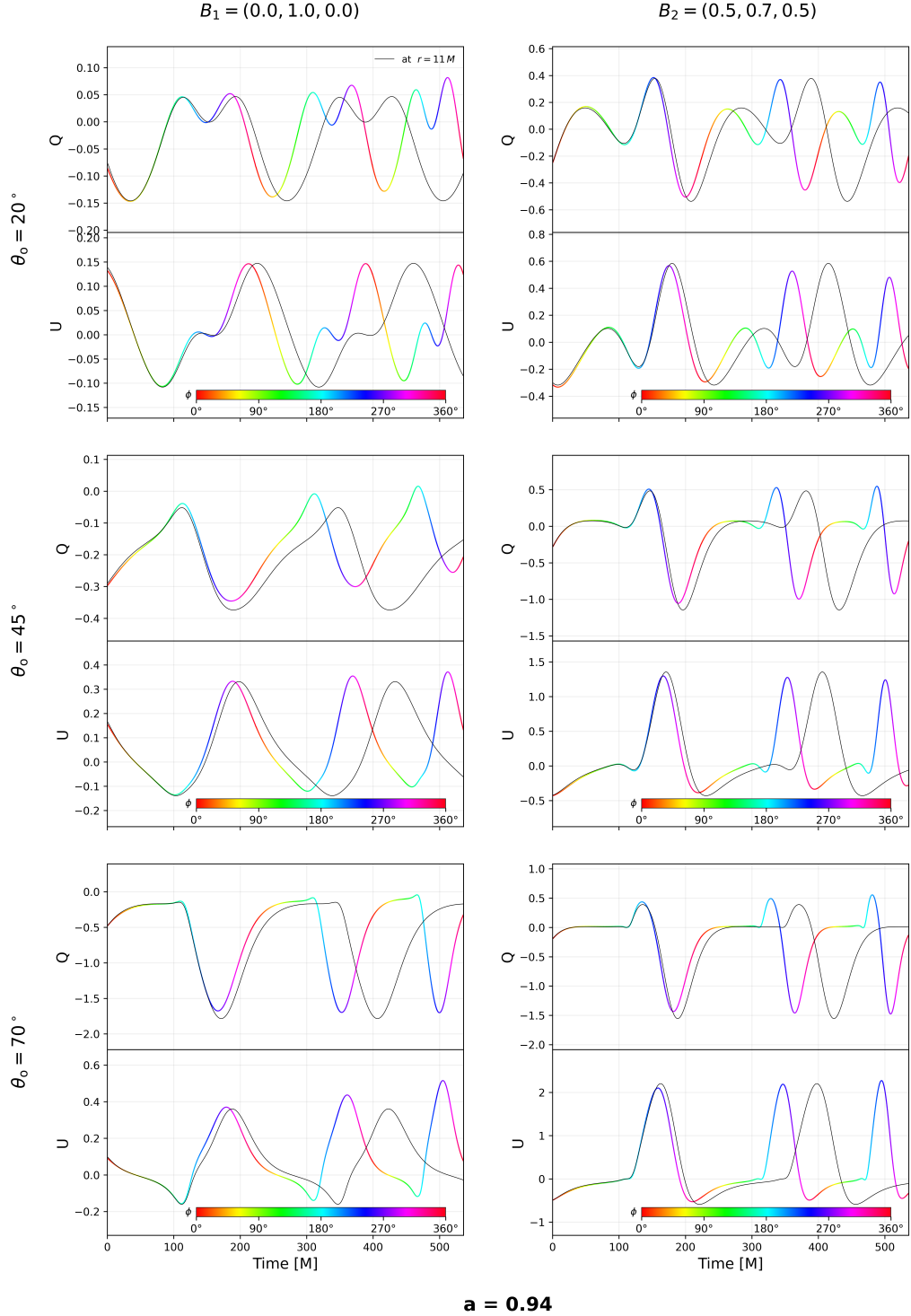}
    \caption{Time evolution of Stokes $Q$ (top row) and $U$ (bottom row) for a hotspot inspiraling into a rapidly spinning black hole ($a=0.94$), which originates at $r=11\,M$. Time is in gravitational units ($M$); for Sgr~A* ($M=4.3\times10^6\,M_\odot$), $100\,M \approx 35$ minutes. Columns vary the magnetic field configuration and the rows the observer's inclination. The colored trajectory (color indicates hotspot's azimuth $\phi$) shows the inspiraling case, while the black line shows, for reference, the stable circular orbit at the initial radius, and \protect\triright{} shows the ISCO crossing. The parameters for the four-velocity (Eq.~\ref{eq:four-velocity}) are $\beta_r = 0.90$ and $\beta_\phi = 0.98$.}
    \label{fig:UvTandQvT_high_spin}
\end{figure*}

The resulting polarimetric signatures produced by inspiraling motion exhibit two notable morphological features. First, the $Q$--$U$ evolution develops secondary, interior loops (nested structure), as shown in Figs.~\ref{fig:low_spin_QU} and~\ref{fig:high_spin_QU}. Second, the loops become elongated as the inclination angle increases. Consistent with these morphological changes, the time series of these Stokes parameters, as shown in Figs.~\ref{fig:UvTandQvT_0spin} and ~\ref{fig:UvTandQvT_high_spin}, are modulated on multiple timescales, with shorter variations associated with orbital motion and longer-timescale evolution as the hotspot nears the ISCO.

Let us now characterize the loop morphology of inspiraling hotspots, taking into account the impact of all the model parameters. First, we will discuss the time-evolving nature of an inspiraling emitter, a distinct characteristic from the perpetual hotspot, where the spin controls the duration of the inspiral and the extent of the multi-loop structure. Then, we will examine the effects on the polarization signature, as we change the parameters of the model: observer's inclination, magnetic field, spectral index, and black hole spin. As we will show, 1) the observer inclination modulates loop amplitude and asymmetry through Doppler boosting and gravitational lensing; 2) the magnetic-field geometry primarily sets the overall loop shape; 3) the spectral index $\alpha_\nu$ controls the contrast between orbital phases through Eq.~\ref{eq:f_obs_scaling}, affecting the apparent completeness and continuity of the loop; and 4) the spin couples to all the other parameters enhancing or augmenting their effects.
We will also comment explicitly on hotspots in the plunging region (inside the ISCO), making contact with recent work Ref.~\cite{Chen:2024jkm} at the end of this Section.

\subsection{Distinct Characteristics of Inspiraling Hotspots}

Inspiraling hotspots exhibit a richer, more intrinsically time-evolving $Q$--$U$ loop morphology than perpetual hotspots. Figure~\ref{fig:trajectories_new} highlights the contrast between inspiral trajectories around a non-rotating ($a=0$) and a rapidly spinning ($a=0.94$) black hole. In the high-spin case, strong frame dragging drives a prolonged, multi-looping trajectory that remains largely tangential until the final plunge. By contrast, in the non-rotating case, the motion develops a dominant radial component, leading to a shorter and more rapid inspiral.

Allowing the hotspot to inspiral also shortens the effective orbital period relative to a stable circular orbit. This behavior is apparent in the time-domain plots (Figs.~\ref{fig:UvTandQvT_0spin} and~\ref{fig:UvTandQvT_high_spin}), where the oscillatory inspiraling signal (colored line) consistently leads in phase relative to the reference circular-orbit signal (black line). This phase lead reflects the evolving orbital dynamics: as the hotspot moves inward, its angular velocity increases.

These dynamical differences translate directly into the complexity of the polarimetric signatures. For example, the Stokes $Q$--$U$ loops for the high-spin case (Fig.~\ref{fig:high_spin_QU}) exhibit more intricate, multi-lobed structures than their low-spin counterparts (Fig.~\ref{fig:low_spin_QU}), particularly when the inspiral proceeds closer to the horizon.

\subsection{Viewing-Angle Dependent Effects: Inclination and Redshift}

The observer’s inclination angle, $\theta_{\rm{o}}$, sets the projection of the orbital motion and modulates the strength of relativistic effects. A key trend is that the amplitude of the $Q$--$U$ loops increases with inclination (from top to bottom in Figs.~\ref{fig:low_spin_QU} and \ref{fig:high_spin_QU}), consistent with stronger Doppler boosting at higher inclinations, which amplifies the polarization signal. For the $\vec{B}_1$ configuration, this amplification is accompanied by a pronounced elongation of the loops.

At high inclinations (e.g., $\theta_{\rm{o}} = 70^\circ$), the orbital velocity has a large component along the line of sight, maximizing the dynamic range of the redshift factor $g$ and producing extreme Doppler modulation. The resulting preference for specific azimuthal angles $\phi$ in the color maps (e.g., $\phi \sim 180^\circ$-- $360^\circ$ in Figs.~\ref{fig:low_spin_QU} and~\ref{fig:high_spin_QU}) follows directly: emission from the receding phase is strongly suppressed by its lower flux. This is also visible in the time-domain signals (Figs.~\ref{fig:UvTandQvT_0spin} and~\ref{fig:UvTandQvT_high_spin}), where increasing inclination produces a plateau following each peak in the emission spectrum; these plateaus correspond to the receding portion of the orbit (e.g., $\phi \sim 0^\circ$-- $180^\circ$).

In many configurations, the combination of large orbital velocities and strong gravity can suppress the receding-phase emission ($\phi \sim 0^\circ$-- $180^\circ$) to the point of near undetectability as $\theta_{\rm{o}}$ increases. In particular, the phases most prominently represented in the $Q$--$U$ loops move to include more emission from the far side, driven by frame dragging in the high-spin case, which alters the portions of the trajectory that have the most favorable kinematics along the photon-emission direction. This effect becomes increasingly important as the trajectory approaches the ISCO, as illustrated in Fig.~\ref{fig:UvTandQvT_high_spin}.

\subsection{The Configuration of the Magnetic Field}
\label{subsec:magnetic_field}

The magnetic-field configuration is the primary determinant of the $Q$--$U$ loop shape and orientation, as is evident from the pronounced differences between the columns in Figs.~\ref{fig:low_spin_QU} and~\ref{fig:high_spin_QU}. For a purely vertical field, $\vec{B}_1 = (0.0,1.0,0.0)$, the low-spin and low-inclination case (Fig.~\ref{fig:low_spin_QU}) shows a clear separation of the hotspot’s azimuthal position $\phi$ into distinct inner and outer tracks in the $Q$--$U$ plane, yielding a direct mapping between orbital phase and polarization state. This separation weakens at higher inclinations, where lensing increasingly dominates the observed signal. When the magnetic field includes significant non-axisymmetric components, e.g., $\vec{B}_2 = (0.5,0.7,0.5)$, this simple mapping breaks down and the loop morphology changes qualitatively. The $Q$--$U$ tracks become shifted, elongated, and otherwise morphologically distinct, demonstrating that the magnetic-field geometry fundamentally sculpts the observed polarization by setting the local direction of the emitted electric vector.

\subsection{The Spectral Index}
\label{subsec:spectral_index}

A critical ingredient, often treated as fixed in studies of perpetual hotspots, is the spectral index $\alpha_\nu$, which governs the detectability of different orbital phases by setting the strength of Doppler boosting. Through its appearance in the observed-flux scaling (Eq.~\ref{eq:f_obs_scaling}), $\alpha_\nu$ directly controls the degree of contrast between the approaching and receding portions of the orbit. Accordingly, the strong bias in the observed azimuthal positions in our results, evident in the color maps of all figures, is a direct consequence of Doppler boosting. The loops shown in Figs.~\ref{fig:low_spin_QU} and \ref{fig:high_spin_QU} are computed assuming $\alpha_\nu = 1$.

Varying $\alpha_\nu$ would quantitatively modify the contrast between the approaching and receding phases. Small values of $\alpha_\nu$ would make the loops appear more complete, and large values of $\alpha_\nu$, more intermittent and dominated by the approaching phase. In this sense, the apparent ``completeness'' of a polarimetric loop is itself informative about the underlying emission spectrum. Future analyses of hotspot flares should therefore treat $\alpha_\nu$ as a key model parameter. In particular, an informative prior on $\alpha_\nu$, ideally guided by simultaneous multiwavelength observations, will be important for breaking degeneracies among orbital dynamics, inclination, and the emission spectrum when interpreting polarimetric data.

\subsection{Plunging Region Hotspots}
\label{subsec:plunging_hotspots}

In this Section, we focus on the emission from hotspots inside the ISCO ($r < r_{\text{ISCO}}$), where the emitting particle plunges toward the event horizon under a generalized, non-geodesic four-velocity prescription. The plunging region has gained significant attention over the past few years in x-ray modeling of sources, due to its potential diagnostic power of accretion flow dynamics~\cite{Wilkins:2020pgu,Cardenas-Avendano:2020xtw}. In the context of hotspots, plunging emitters were studied in Ref.~\cite{Chen:2024jkm} using Cunningham’s~\cite{Cunningham:1975zz} geodesic model and two specific magnetic-field prescriptions: (i) a vertical field described by the source-free Papapetrou--Wald solution, and (ii) a radial field sourced by an accretion flow infalling from infinity under ideal magnetohydrodynamic assumptions. Therein, approximate analytic lensing expressions for plunging geodesics and homoclinic orbits were also derived for an observer located on-axis. As we show below, allowing for non-Keplerian motion enhances the phenomenology described in the previous Sections for inspiral motion outside the ISCO.

The time-domain evolution, shown in Figs.~\ref{fig:UvTandQvT_0spin} and~\ref{fig:UvTandQvT_high_spin}, reveals a sharp contrast in the signal’s decay between the two spin cases. For low spin, the polarized flux, $Q(t)$ and $U(t)$, terminates abruptly as the hotspot trajectory becomes nearly radial, leading to rapid gravitational redshift and to lensing that removes much of the emission from the observer’s screen. For high spin, the signal instead resembles a damped oscillator decaying smoothly over several orbits of the prolonged, largely tangential inspiral. In this case, distinct plateaus follow each modulation cusp in $Q(t)$ and $U(t)$, corresponding to phases when the hotspot azimuth $\phi$ is in the approaching portion of the orbit (between $180^\circ$ and $360^\circ$), where Doppler boosting maximizes the observed signal. This enhancement becomes stronger with the increasing observer's inclination $\theta_{\rm o}$. More generally, the extended inspiral around a high-spin black hole allows the hotspot to sample a broader range of orbital phases and gravitational redshifts, producing these more structured polarization patterns.

\begin{figure*}[!ht]
    \centering
    \includegraphics[width=0.75\textwidth]{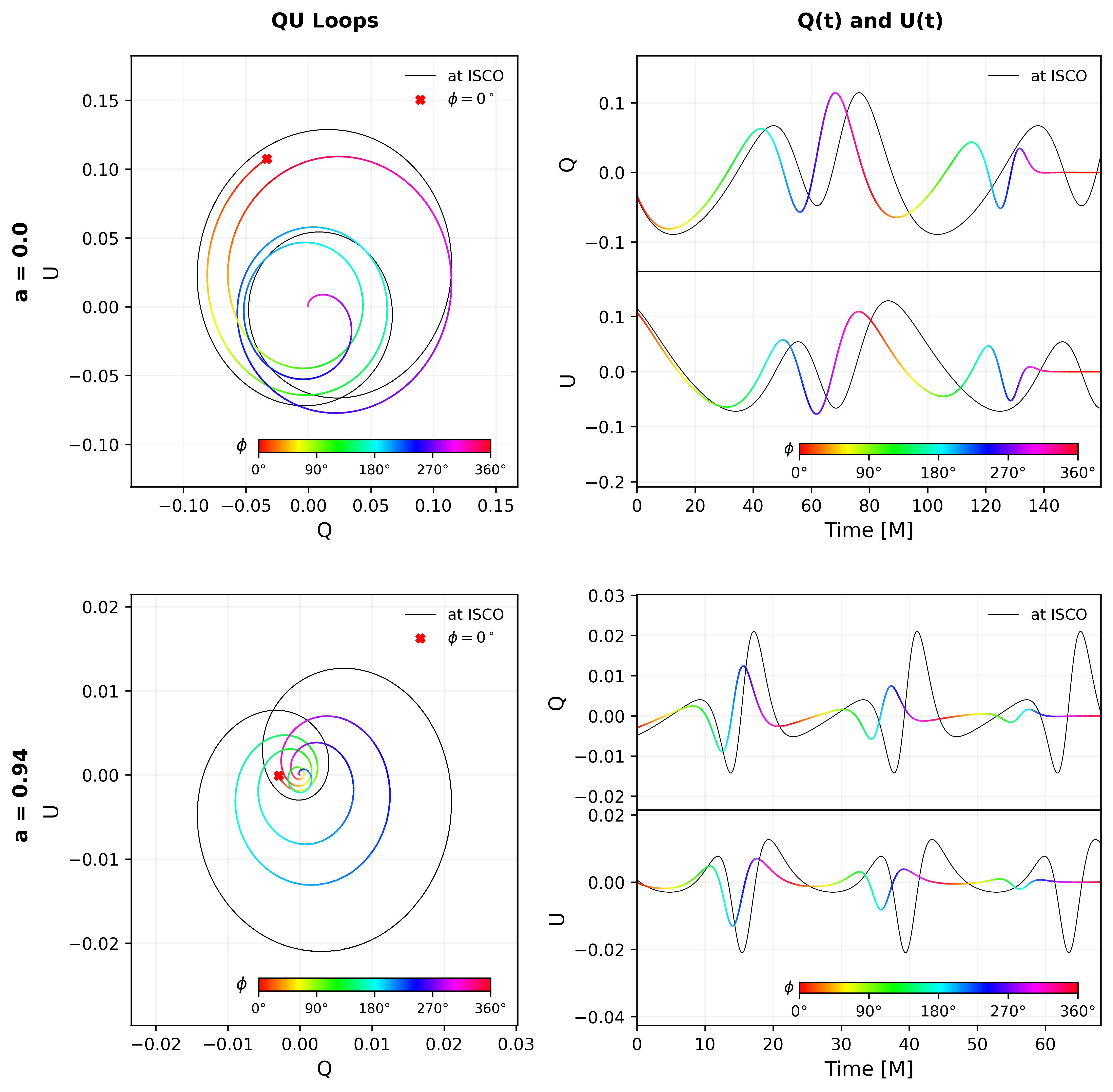}
    \caption{Polarimetric $Q$--$U$ loops (left column) for a hotspot plunging toward the event horizon around a non-spinning black hole ($a=0$, top left) and a rapidly spinning black hole ($a=0.94$, bottom left). The right column shows the corresponding time variability for each case. All panels are computed at low inclination ($\theta_{\rm{o}}=20^\circ$) with a purely vertical magnetic field configuration ($\vec{B}_1=(0.0,1.0,0.0)$). The colored trajectory encodes the instantaneous hotspot azimuth $\phi$, while the black curve denotes the resulting $Q$--$U$ loop generated by a hotspot at a stable circular orbit at the ISCO for reference. The four-velocity parameters are set to $\xi = \beta_r = \beta_\phi = 1$, thus, recovering Cunningham's disk model for the plunging region. For these cases, the plunge is initialized at $0.95\,r_{\rm ISCO}$ (because a stable orbit at the ISCO would not plunge) and followed down to $1.005\,r_H$, where $r_H$ is the horizon radius.}
    \label{fig:Cunningham_plunging_QU}
\end{figure*}

Figure~\ref{fig:Cunningham_plunging_QU} shows the $Q$--$U$ loops obtained using Cunningham’s four-velocity prescription, as in Ref.~\cite{Chen:2024jkm}, but evaluated for the same magnetic-field configurations adopted throughout this work (which differ from those considered in Ref.~\cite{Chen:2024jkm}). This provides a convenient baseline---which we recover by setting $\xi=1$ and $\beta_r=\beta_\phi=1$---for isolating the impact of our more general velocity prescription. Relative to Cunningham’s geodesic plunge, the parametric four-velocity can substantially modify the trajectory through the ``Keplerianity'' parameters, thereby reshaping the resulting $Q$--$U$ morphology. In the plunging region, these deviations can accelerate the inspiral and enhance the systematic drift of the polarimetric centroid, leading to more pronounced shifts in the loop evolution.

Within Cunningham’s model, a hotspot initiated at the ISCO remains on a stable orbit and therefore does not plunge. Accordingly, for Fig.~\ref{fig:Cunningham_plunging_QU} we initialize the hotspot at $0.95\,r_{\rm ISCO}$ and follow it down to $1.005\,r_H$, where $r_H$ is the horizon radius. While the two cases exhibit broadly different qualitative features, the most apparent distinction between Figs.~\ref{fig:UvTandQvT_0spin} and~\ref{fig:UvTandQvT_high_spin} and Fig.~\ref{fig:Cunningham_plunging_QU} is the total time required to reach the horizon, as seen in the corresponding time-domain variability. This highlights an important aspect of the non-geodesic model: even seemingly small departures from Cunningham’s prescription~\cite{Cunningham:1975zz} can lead to substantially different dynamics.

\section{Discussion}
\label{sec:discussion}

In this work, we revisited the anatomy of $Q$--$U$ loops and computed them, for the first time, for inspiraling hotspots outside the ISCO. Compared to perpetual hotspots, i.e., hotspots on stable orbits at constant radius, inspiraling hotspots naturally produce a richer, time-dependent morphology while still exhibiting the characteristic loop-like behavior on orbital timescales. This modeling is especially relevant given that previous works have shown that perpetual hotspots can only account for an initial primary loop, whereas additional data points often fall inside the primary loop, a morphology that inspiraling trajectories can naturally generate. 

The examples presented here reveal a clear hierarchy in how model parameters shape the polarimetric signatures. The magnetic-field geometry is the dominant factor setting the overall $Q$--$U$ loop shape. The observer's inclination, $\theta_{\rm{o}}$, modulates the loop’s asymmetry and amplitude by controlling the relative importance of Doppler boosting and gravitational lensing. The spectral index, $\alpha_\nu$, often treated as fixed, regulates the contrast between orbital phases and therefore affects the apparent completeness, size, and continuity of the loop; it should be treated as a free parameter in hotspot modeling. Lastly, the black hole spin governs the characteristic dynamical timescales and the extent of the multi-loop structure during the terminal stages of the inspiral, as reflected in the duration and complexity of the loop's features as the orbit approaches and crosses the ISCO. Although spin is not the dominant parameter that defines the overall loop morphology, it and can modulate the signatures induced by the other parameters. Because the detailed $Q$--$U$ morphology encodes information about the spin and the magnetic-field structure, this framework can be refined to extract such constraints directly from future polarimetric data.

In this work, for the examples presented, we assume that the hotspot survives throughout the plunge, up to the point where redshift effectively suppresses its emission. While it represents a step toward a more general description of $Q$--$U$ loops produced by hotspots, it still neglects many additional considerations, which may be important for modeling realistic observations. In practice, the hotspot may not remain coherent along the entire trajectory, which would truncate the loop and limit the observable phases.  We also neglect the hotspot’s finite spatial extent~\cite{Broderick:2005jj,Zamaninasab:2009df,Vos:2022yij}; during an inspiral, tidal effects may deform the emitting region and thereby modify the observed polarization. We further restrict attention to equatorial motion, although out-of-plane trajectories may also be relevant~\cite{GRAVITY:2020lpa}. Finally, we consider only the direct-image contribution; as discussed in Ref.~\cite{Wielgus:2022heh}, higher-order images are expected to become increasingly important as the hotspot approaches the black hole. Incorporating these effects is beyond the scope of this work and is left for future studies.

While the Kerr metric remains the prevailing model for the spacetime of Sgr~A*, recent and future horizon-scale observations provide opportunities to explore potential deviations from general relativity~\cite{EventHorizonTelescope:2022wkp,GRAVITY:2023avo,Lupsasca:2024xhq}. Although current constraints suggest that uncertainties associated with the spacetime geometry are likely subdominant to astrophysical modeling uncertainties~\cite{Bauer:2021atk,Cardenas-Avendano:2023obg}, continued improvements in data quality and in modeling of the emission physics may make such tests increasingly feasible. In several modified theories of gravity, the characteristic radii can shift, and the critical dynamical threshold at which stable orbital motion transitions into a rapid, non-periodic plunge may be significantly altered. Consequently, modeling infalling hotspots in modified theories of gravity~\cite{Kocherlakota:2024hyq,Rosa:2025pqp} can provide a means to probe whether the stable-to-plunging transition aligns with Kerr predictions.\\

\acknowledgments

We thank Z.~Gelles, E.~Levati, M.~Wielgus, P.~Wu and H.~Zhu for clarifying conversations and valuable comments on this project. We also thank an anonymous referee for their constructive comments. Computations were performed using the Wake Forest University (WFU) High Performance Computing Facility, a centrally managed computational resource available to WFU researchers including faculty, staff, students, and collaborators~\cite{WakeHPC}. 
D.G. acknowledges support from the Harvard Postdoctoral Fellowship for Future Faculty Leaders, the National Science Foundation (AST-2307887), the Gordon and Betty Moore Foundation (Grant \#8273.01), and the John Templeton Foundation (Grant \#62286). The opinions expressed in this publication are those of the authors and do not necessarily reflect the views of these Foundations.

\section*{DATA AVAILABILITY}
The code used to produce the results of this study is openly available in the Zenodo repository at:~\href{https://doi.org/10.5281/zenodo.20060519}{https://doi.org/10.5281/zenodo.20060519}~\cite{pablomartin_2026_20060519}.

\section*{References}

\bibliographystyle{utphys}
\bibliography{refs}

\end{document}